\documentclass{article}
\pdfoutput=1
\usepackage{jcappub}

\title{The Bispectrum of $f(R)$ Cosmologies}
\author[a,b]{H\'ector Gil-Mar\'in,}
\author[c]{Fabian Schmidt,}
\author[d]{Wayne Hu,}
\author[e,b]{Raul Jimenez,}
\author[e,b]{ and Licia Verde}

\affiliation[a]{Institute of Space Sciences (IEEC-CSIC), Faculty of Science, Campus UAB, Bellaterra 08193, Spain}
\affiliation[b]{Institute of Sciences of the Cosmos (ICC-IEEC), University of Barcelona, Barcelona 08024, Spain}
\affiliation[c]{Theoretical Astrophysics, California Institute of Technology, Mail Code 350-17, Pasadena, California 91125}
\affiliation[d]{Kavli Institute for Cosmological Physics, Department of Astronomy \& Astrophysics, University of Chicago, Chicago, IL 60637}
\affiliation[e]{ICREA Instituci\'o Catalana de Recerca i Estudis Avan\c{c}ats}

\emailAdd{gil@ieec.uab.es}
\emailAdd{fabians@caltech.edu}
\emailAdd{whu@background.uchicago.edu}
\emailAdd{raul.jimenez@icc.ub.edu}
\emailAdd{liciaverde@icc.ub.edu}

\abstract{In this paper we analyze a suite of cosmological simulations of modified gravitational action $f(R)$ models, where cosmic acceleration is induced by a scalar field that acts as a fifth force on all forms of matter. In particular, we focus on the bispectrum of the dark matter density field on mildly non-linear scales.   For models with the same {\it initial} power spectrum, the dark matter  bispectrum shows significant differences for cases where the final dark matter power spectrum also differs.  
Given the different dependence on bias of the galaxy power spectrum and bispectrum, bispectrum measurements can close the loophole of galaxy bias hiding differences in the power spectrum.   Alternatively, changes in the initial power spectrum can also hide differences.   By constructing $\lcdm$ models with very similar  {\it final} non-linear power spectra, we show that the differences in the  bispectrum are reduced ($\lesssim4\%$) and are comparable with differences in the imperfectly matched power spectra.  These results indicate that the bispectrum depends mainly on the power spectrum and less sensitively on the gravitational signatures of the $f(R)$ model.   This weak dependence of the matter bispectrum on gravity makes it useful for breaking degeneracies associated with galaxy bias, even for models beyond general relativity.|}

\begin{document}

\newcommand{\eff}{{\rm eff}}
\newcommand{\lcdm}{\Lambda{\rm CDM}}
% Subscript to denote matched LCDM simulations.  Denote however we want.
\newcommand{\matched}{{\rm matched}}
\newcommand{\fR}{$f(R)$\,\,\,}
\maketitle

%\begin{keywords}
%cosmology: theory - cosmology: cosmological parameters - cosmology: large-scale structure of Universe
%\end{keywords}

%------------------------------------------------------------------------------------------------------------

\section{Introduction}

Observations of Type Ia supernovae suggest that the Universe has been accelerating since redshift $z\sim0.5$ \citep{acceleration1,acceleration2}. Today the physical mechanism responsible for this process is still a mystery.  
The simplest model to explain the acceleration of the Universe is the $\Lambda$CDM (Lambda Cold Dark Matter model). This model assumes that the  acceleration is driven by an exotic form of energy with negative pressure that might be related to the vacuum energy of quantum field theories. This theory is equivalent to adding an integration constant to the Einstein equations. 

Alternative theories to the vacuum energy propose a modification of gravity in the infrared that would produce an accelerated expansion. One possibility are the $f(R)$ class of models (see \cite{fr} and references therein). These models produce accelerated expansion th\-rough a modification of the Einstein-Hilbert action by an arbitrary function of the Ricci scalar $R$. As a consequence, an extra propagating scalar field  appears that mediates  a fifth force on all forms of matter. The range of this force depends on the functional form of $f(R)$. In order to satisfy solar system tests, $f(R)$ models are often chosen to present a chameleon behavior. The chameleon mechanism makes the extra scalar field become increasingly massive in higher-curvature regions, suppressing the range of the fifth force in dense environments. 

In previous works, cosmological simulations \citep{I} have been used to study the power spectrum \citep{II} and halo statistics \citep{III} of these kinds of models. More recent studies with higher resolution have confirmed these previous results \citep{ZhaoEtal} and extended the investigation to smaller scales. In the present work we focus on how the dark matter bispectrum is modified in this class of models. While these models also predict a non-linear matter power spectrum different from the $\Lambda$CDM one, it is nevertheless interesting to look at the bispectrum for at least two reasons: {\it a)} except for gravitational lensing, measurements of clustering yield the galaxy or the baryon power spectrum, not the dark matter one: as baryonic physics and galaxy formation are complicated phenomena, the observed power spectrum may  be biased, i.e. may differ significantly from the dark matter one;  the bispectrum is well known  for helping disentangle effects of gravity from effect of biasing e.g.,  \cite{fry94,verdeetal02}. {\it b)} once we allow ourselves to consider non-standard models,  the initial (linear) matter power spectrum does not have to be the power-law $\Lambda$CDM one to reproduce the observations. The form of the bispectrum kernel is a possible ``signature" of gravity as it gets modified by any modifications from GR behavior e.g., \cite{sealfon}.

Here we pay special attention to see whether the bispectrum can be used to break degeneracies between models with the same observed power spectrum and the same cosmology, but different gravity.
We begin in \S\ref{dynamics} with a review of non-linear gravitational dynamics in $f(R)$ models, in \S\ref{simulations} we briefly describe the simulations and in \S\ref{bispectrum} we introduce the density field statistics. We discuss the results in \S\ref{results} and conclude in \S\ref{conclusions}.

\section{\fR Gravity}\label{dynamics}
The $f(R)$ class of models generalizes the Einstein-Hilbert action to include a function $f(R)$ of the Ricci scalar $R$,
\begin{equation}
 S=\int d^4x\, \sqrt{-g}\left[\frac{R+f(R)}{16\pi G}+L_m\right]\,.
\label{action}
\end{equation}
Here $L_m$ is the Lagrangian of matter and we have assumed $c=\hbar=1$. For standard GR with a cosmological constant, $f(R)=-16\pi G\rho_\Lambda$, whereas for modified gravity, the force modification is associated with an additional scalar degree of freedom $f_R\equiv df/dR$. In particular, in this paper we use the model for $f(R)$ proposed by \cite{hu_sawicki},
\begin{equation}
 f(R)\propto \frac{R}{AR+1}\,,
\end{equation}
where $A$ is a constant with dimensions of length squared. We can write this equation as a function of its derivative evaluated at $\bar R_0$ (the background curvature today), namely $f_{R0}$. We adjust the proportionality constant to match some effective cosmological constant $\rho_\Lambda$ in the limit where $f_{R0}\rightarrow0$. For high enough curvature such that $AR\gg1$, $f(R)$ can then be approximated as,
\begin{equation}
 f(R)=-16\pi G \rho_\Lambda-f_{R0}\frac{\bar R_0^2}{R}.
\end{equation}
The modified Einstein equations can be computed by varying the Einstein-Hilbert action (Eq. \ref{action}) with respect to the metric. We work in the quasistatic limit where the time derivatives are negligible compared to the spatial derivatives. In this regime, valid on scales much smaller than the horizon $1/H$, the trace of the modified Einstein equations yields the $f_R$ field equation,
\begin{equation}
 \label{nabla2_df}\nabla^2 \delta f_R=\frac{a^2}{3}\left[\delta R(f_R)-8\pi G\delta \rho_m\right]\,,
\end{equation}
where $a$ is the scale factor, $\delta f_R=f_R(R)-f_R(\bar R)$, $\delta R=R-\bar R$ and $\delta\rho_m=\rho_m-\bar\rho_m$. Here $\bar R$ is the background curvature that can be approximated by a $\Lambda$CDM universe for $|f_{R0}|\ll 1$ and $\rho_m$ ($\bar\rho_m$) is the (background) matter density.

On the other hand, the time-time component of the Einstein equations yields the modified Poisson equation,
\begin{equation}
\label{poisson} \nabla^2 \Psi=\frac{16\pi G}{3}a^2\delta\rho_m-\frac{a^2}{6}\delta R(f_R)
\end{equation}
where  $\Psi = \delta g_{00}/(2 g_{00})$ is the Newtonian potential. 

For small fluctuations of the field, we can approximate $\delta R\simeq(dR/df_R)|_{\bar R}\delta f_R$. We will refer to this linearization as the non-chameleon limit. Conversely if the field fluctuations are large enough such that $\delta R(f_R)$ cannot be linearized,  the chameleon mechanism operates.  We will refer to use of the exact, as opposed to linearized, equations as full $f(R)$ or just chameleon models.

The linearized field equations formed by Eqs. \ref{nabla2_df} - \ref{poisson} can be solved for the Newtonian potential as a function of the density field. In the linear approximation for $\delta R$, these two equations in Fourier space yield, 
\begin{equation}
\label{poisson_fourier} k^2\Psi({\bf k})=-4\pi G\left(\frac{4}{3}-\frac{1}{3}\frac{\bar\mu^2a^2}{k^2+\bar\mu^2a^2}\right)a^2\delta\rho_m({\bf k}).
\end{equation}
This equation is identical to the one in GR but with a modification of the gravitational constant, 
\begin{equation}
 G_{\rm eff}({\bf k},t)\equiv G\left(\frac{4}{3}-\frac{1}{3}\frac{\bar\mu(t)^2a(t)^2}{k^2+\bar\mu(t)^2a(t)^2}\right).
 \label{Geff}
\end{equation}
Here $\mu(R)\equiv(3df_R/dR)^{-1/2}$ is the effective mass of the scalar field $f_R$ and $\bar\mu$ just stands for $\mu(\bar R)$. The dependence on time is introduced though $\bar R(t)$.
Note that when $f_R\rightarrow0$, $G_{\rm eff}\rightarrow G$ and we recover the $\Lambda$CDM limit, as expected.
It is interesting to see that for a given value of $f_{R0}$ there are two different regimes for $G_{\rm eff}$, depending on whether the physical scale we are studying is larger or smaller than the inverse mass of the field. On large scales $k\ll\mu(t) a(t)$, $G_{\rm eff}\rightarrow G$ and gravity behaves as GR, whereas on small scales $k\gg\mu(t) a(t)$ $G_{\rm eff}\rightarrow 4G/3$ and gravity is stronger than in GR by a factor of $4/3$. 

In other words, in Eq. \ref{poisson_fourier}, one assumes that the mass of the scalar field $\mu$ only depends on time and is the same in all regions of the Universe at a given epoch. However, for cosmologically interesting values of $\mu$ the field is then essentially massless within the Solar System.  The presence of such a scalar field (fifth force) is ruled out by light deflection and time delay measurements in the Solar System, which are all consistent with GR.  
In the full non-linear $f(R)$ theory, $R \propto f_R^{-1/2}$ can become very large in dense environments, suppressing the field and restoring the GR relation $\delta R = 8\pi G \delta\rho_m$ (Eq.~\ref{nabla2_df}).  Thus, gravity is not modified in the same way everywhere, but depends on environment. In regions with large potential wells (inside halos) the mass of the scalar field becomes large and therefore the effective range of interaction of this field shrinks recovering GR. We call this the chameleon mechanism.

\section{Simulations}\label{simulations}
The simulations used in this paper are described in previous works \citep{I,II,III}.  Briefly, the field equation for $f_R$ (Eq.~\ref{nabla2_df}) is solved on a regular grid using relaxation techniques and multigrid iteration.  
The potential $\Psi$ is computed from the density and $f_R$ fields following Eq.~\ref{poisson} using the fast Fourier transform method. The dark matter particles are then moved according to the gradient of the computed potential, $-\nabla \Psi$, using a second order accurate leap-frog integrator.

The simulations were run using the values of $|f_{R0}|=10^{-4}$, $10^{-5}$, $10^{-6}$ (both for chameleon and non-chameleon cases) and 0, which is equivalent to $\Lambda$CDM\footnote{In this paper we are always using negative values for $f_{R0}$, so when we talk about $f_{R0}$ we refer to its absolute value.}. The background expansion history for all cases  differ from $\Lambda$CDM only at ${\cal O}(f_{R0})$ and are hence practically indistinguishable. The cosmology used is $\Omega_\Lambda=0.76$, $\Omega_m=0.24$, $\Omega_b=0.04181$, $H_0=73\,\mbox{km/s/Mpc}$ and initial power in curvature fluctuations $A_s=(4.89\times10^{-5})^2$ at $k=0.05\,\mbox{Mpc}^{-1}$ with a tilt of $n_s=0.958$. This initial power spectrum does not include the effects of baryon acoustic oscillations. Specifically, the initial conditions for the simulations were created using ENZO \citep{enzo}, a publicly available cosmological N-body + hydrodynamics code. ENZO uses the Zel'dovich approximation to displace particles on a uniform grid according to the initial power spectrum. In order to propagate the initial power spectrum until late times the transfer function from \citet{EH} was used. 

The simulations were started at $a=0.02$ and are integrated in time in steps of $\Delta a=0.002$.
All simulations used here correspond to boxes of comoving size $L=256$ and $400$ $\mbox{Mpc/}h$ with $512^3$ grid cells and $256^3$ particles. For each box size, we have 6 runs for each value of $f_{R0}$, with different realizations of the initial conditions.

\section{Power Spectrum and Bispectrum}\label{bispectrum}

The simplest statistic of interest of the matter density field is the power spectrum $P(k)$, defined by the second moment of the Fourier amplitude of the density contrast,
\begin{equation}
 \langle \delta({\bf k})\delta({\bf k'})\rangle \equiv (2\pi)^3 \delta^D({\bf k+k'}) P(k)\,,
\label{Pk}
\end{equation}
where $\langle\dots \rangle$ denotes the ensemble average over different realizations of the Universe. By statistical isotropy, the power spectrum does not depend on the direction of the ${\bf k}$-vector. In practice we only have one observable Universe, so the average $\langle\dots \rangle$ cannot be computed. However, using the isotropy of the power spectrum we can compute the average over all different directions for each $\bf k$-vector.
Note also that $P(k)$ is defined to be real. Since ${\bf k}=-{\bf k'}$, $\delta({\bf k})\delta({\bf k'}) \sim |\delta({\bf k})|^2$, which is a real number.

The second statistic of interest is the bispectrum $B$, defined by,
\begin{equation}
 \langle \delta({\bf k}_1) \delta({\bf k}_2) \delta({\bf k}_3)\rangle\equiv (2\pi)^3 \delta^D({\bf k}_1+{\bf k}_2+{\bf k}_3) B({\bf k}_1, {\bf k}_2, {\bf k}_3).
\label{Bk}
\end{equation}
The Dirac delta function $\delta^D$, ensures that the bispectrum is defined only for ${\bf k}$-vector configurations that form closed triangles: $\sum_i {\bf k}_i=0$. Note that once the average is taken, the imaginary part of $\delta({\bf k}_1) \delta({\bf k}_2) \delta({\bf k}_3)$ goes to zero.

It is convenient to define the reduced bispectrum $Q_{123}\equiv Q({\bf k}_1, {\bf k}_2, \bf{k}_3)$ as,
\begin{equation}
Q_{123}\equiv \frac{B({\bf k}_1, {\bf k}_2, {\bf k}_3)}{P(k_1)P(k_2)+P(k_1)P(k_3)+P(k_2)P(k_3)},
\end{equation}
which takes away most of the dependence on scale and cosmology. The reduced bispectrum is useful when comparing different models, since it has a weak dependence on cosmology and one can thus break degeneracies between cosmological parameters to isolate the effects of gravity. Hereafter, when we speak of the bispectrum we are always referring to the reduced bispectrum.

\section{Results}\label{results}
In this paper, we present two ways of comparing the $f(R)$ and $\Lambda$CDM reduced bispectra we obtain from N-body simulations. 
The differences depend on whether the models are matched in their initial or final power
spectra.
In method A we compare the output bispectra from N-body simulations with the same initial power spectra.  Thus some of the difference in the bispectra can be attributed to the different amounts of final nonlinear power in the two sets.  
Method B tries to separate these contributions by generating modified initial power spectrum $\Lambda$CDM simulations whose power spectra at $z=0$ match those of the $f(R)$ simulations.  

For both methods we compute the bispectrum randomly drawing $k$-vectors from a specified bin, namely $\Delta k$ and randomly orientating the triangle in space. We make the number of random triangles to depend on the number of fundamental triangle per bin, that scales as $k_1 k_2 k_3 \Delta k^3$ \citep{SC97}. In this paper we always choose $\Delta k=3k_{min}$. For the equilateral case, at scales of  $k\sim0.65$ $h$/Mpc we are generating $\sim5\times10^8$ triangles. We have verified that increasing the number of triangles beyond this value does not have any effect on the measurement.

\subsection{Method A (matched initial power spectrum)}\label{methodA}

Our first test of $f(R)$ vs $\Lambda$CDM bispectra utilize the same initial power spectrum.   
In our $f(R)$ models, modifications to gravity go to zero rapidly with redshift and the
expansion history differs negligibly from $\Lambda$CDM.   Thus models with the same initial power spectra as $\Lambda$CDM fit observations at high redshift, such as primary CMB anisotropy, equally well.

Since all N-body simulations start from the same initial power spectrum, the $f(R)$ modifications to gravity during the acceleration epoch lead to differences in the
dark matter power spectra at low redshift that increase with $|f_{R0}|$
as  was noted in Fig. 2 of \cite{II}: these differences reach up to $\sim50\%$ for $|f_{R0}|=10^{-4}$ and $\sim10\%$ for $|f_{R0}|=10^{-6}$ for $k\simeq0.5\,h/\mbox{Mpc}$ with respect to the $\Lambda$CDM model. 

Bispectra for matched initial power spectra but differing final power spectra is what is usually computed analytically \citep{bernardeau,borisov}: one predicts (using the modified Euler and continuity equations in perturbation theory) the reduced bispectra for different gravity models starting from a given initial $\delta_k$ field.  One might expect that the reduced bispectra differences are independent of the power spectra, but this is only strictly true for equilateral configuration and only in the tree-level regime (for $k<0.06$ $h$/Mpc at $z=0$). This is the main caveat of method A: the differences seen in the reduced bispectrum could be due to differences in the final matter power spectrum and not unique
signatures of $f(R)$ gravity.

\begin{figure}
 \centering
\includegraphics[scale=0.2]{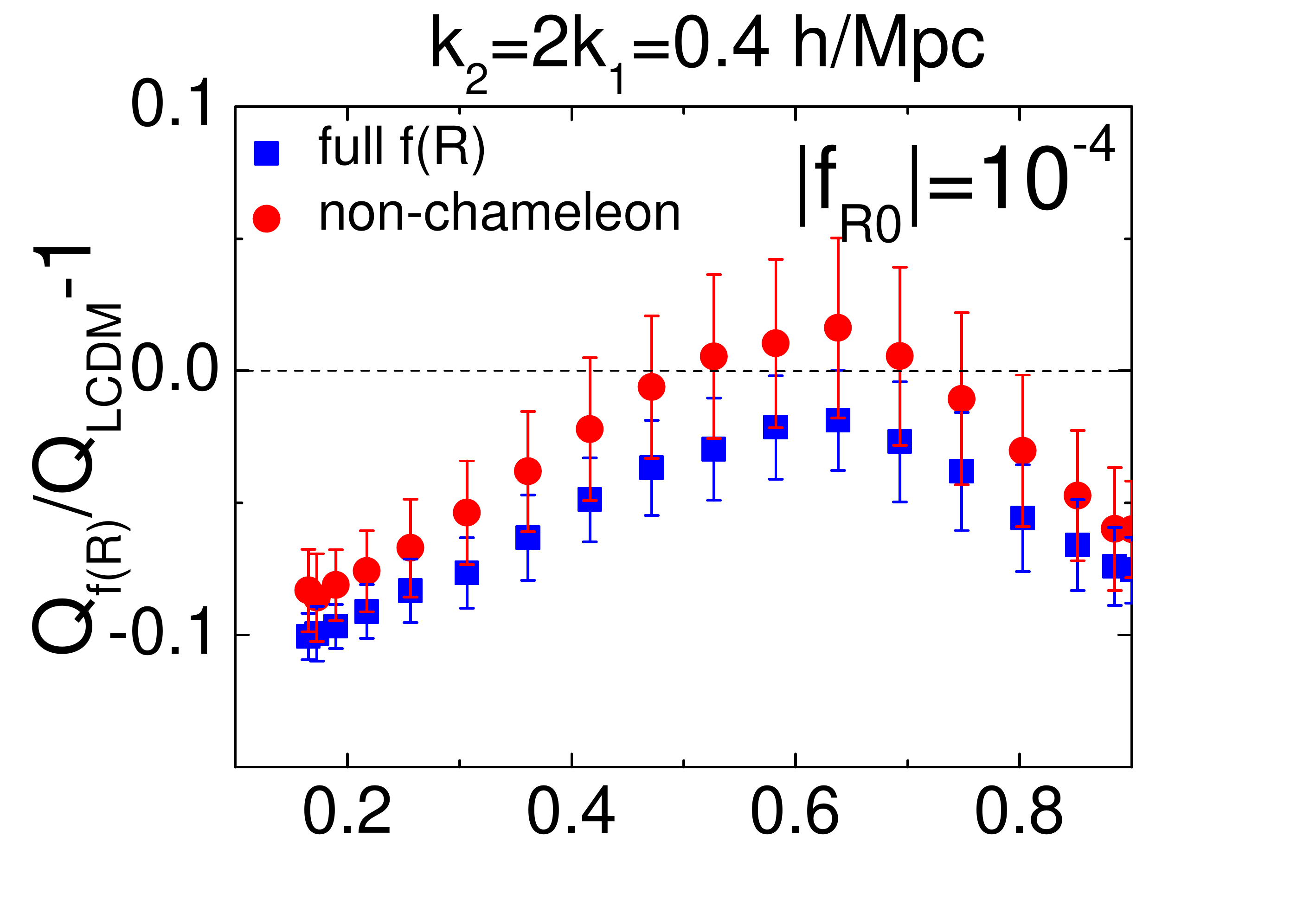}
\includegraphics[scale=0.2]{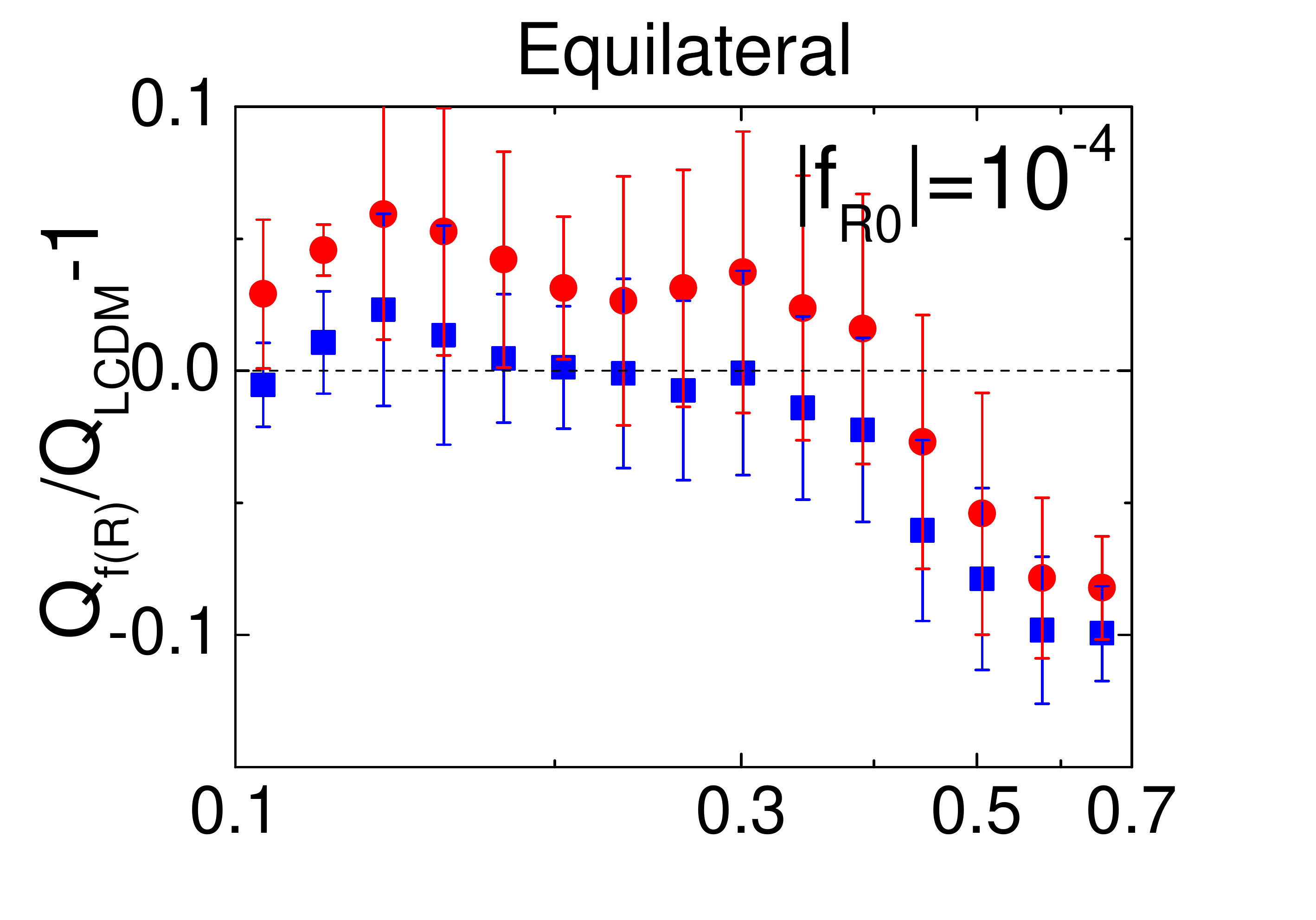}

\includegraphics[scale=0.2]{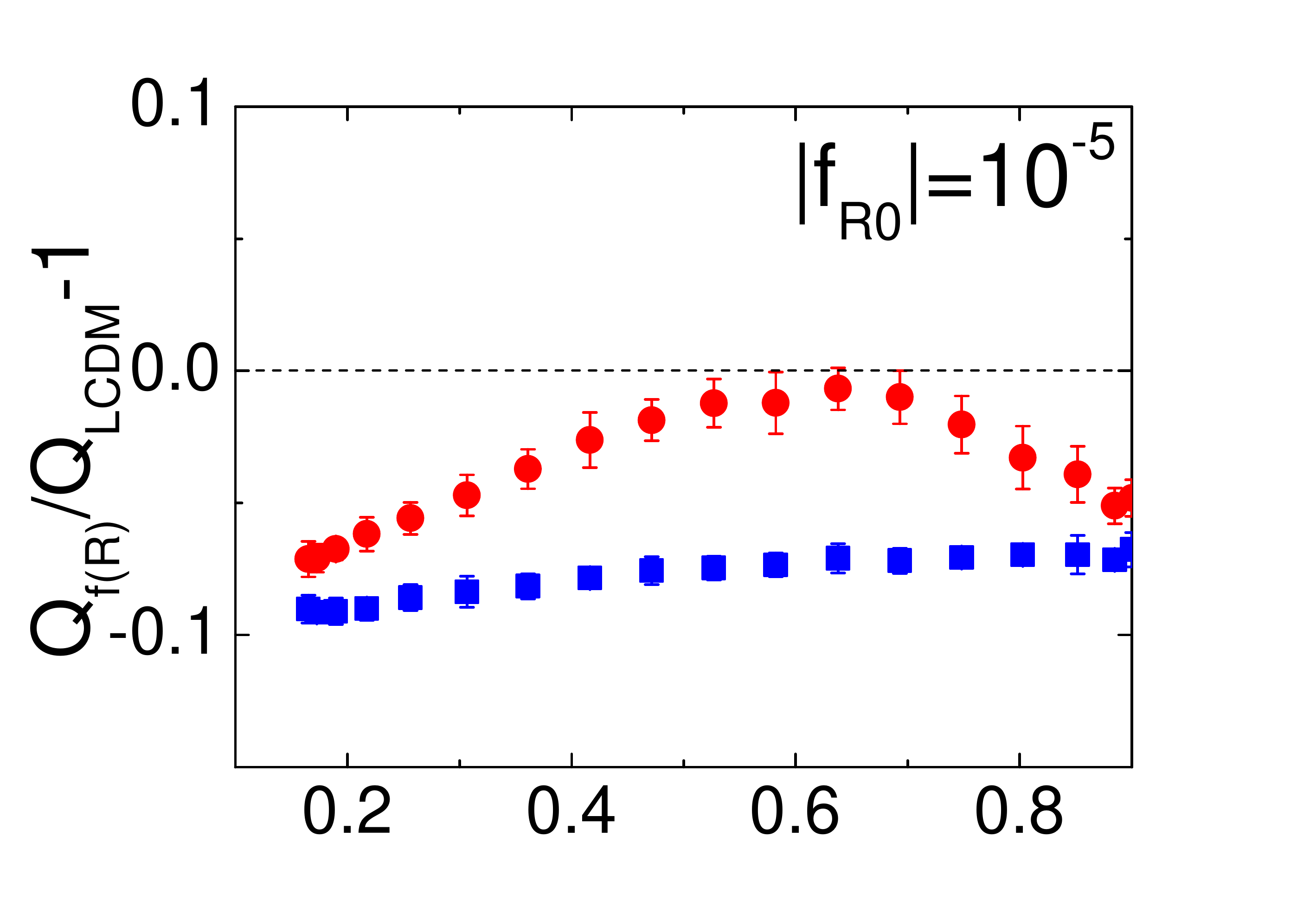}
\includegraphics[scale=0.2]{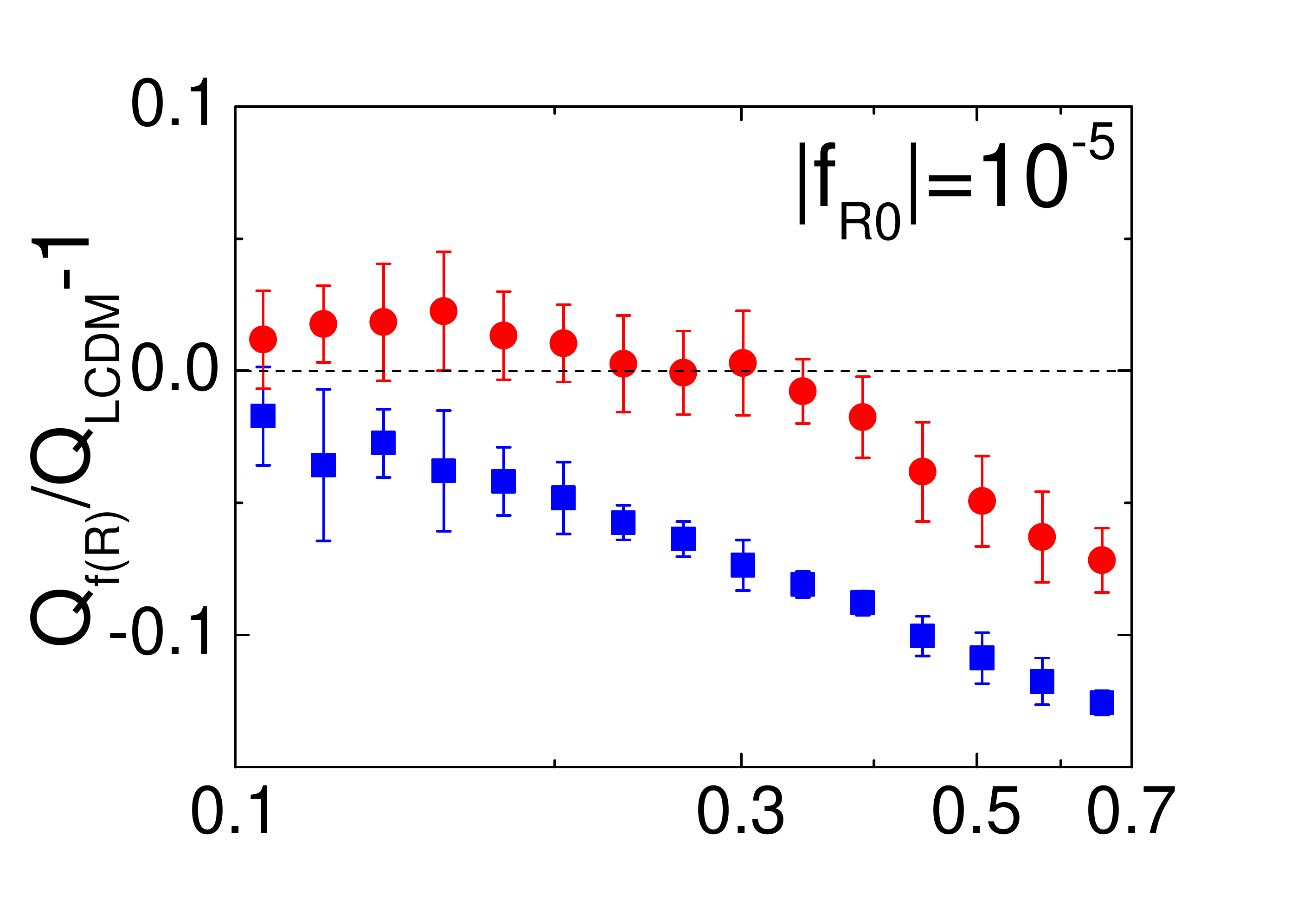}

\includegraphics[scale=0.2]{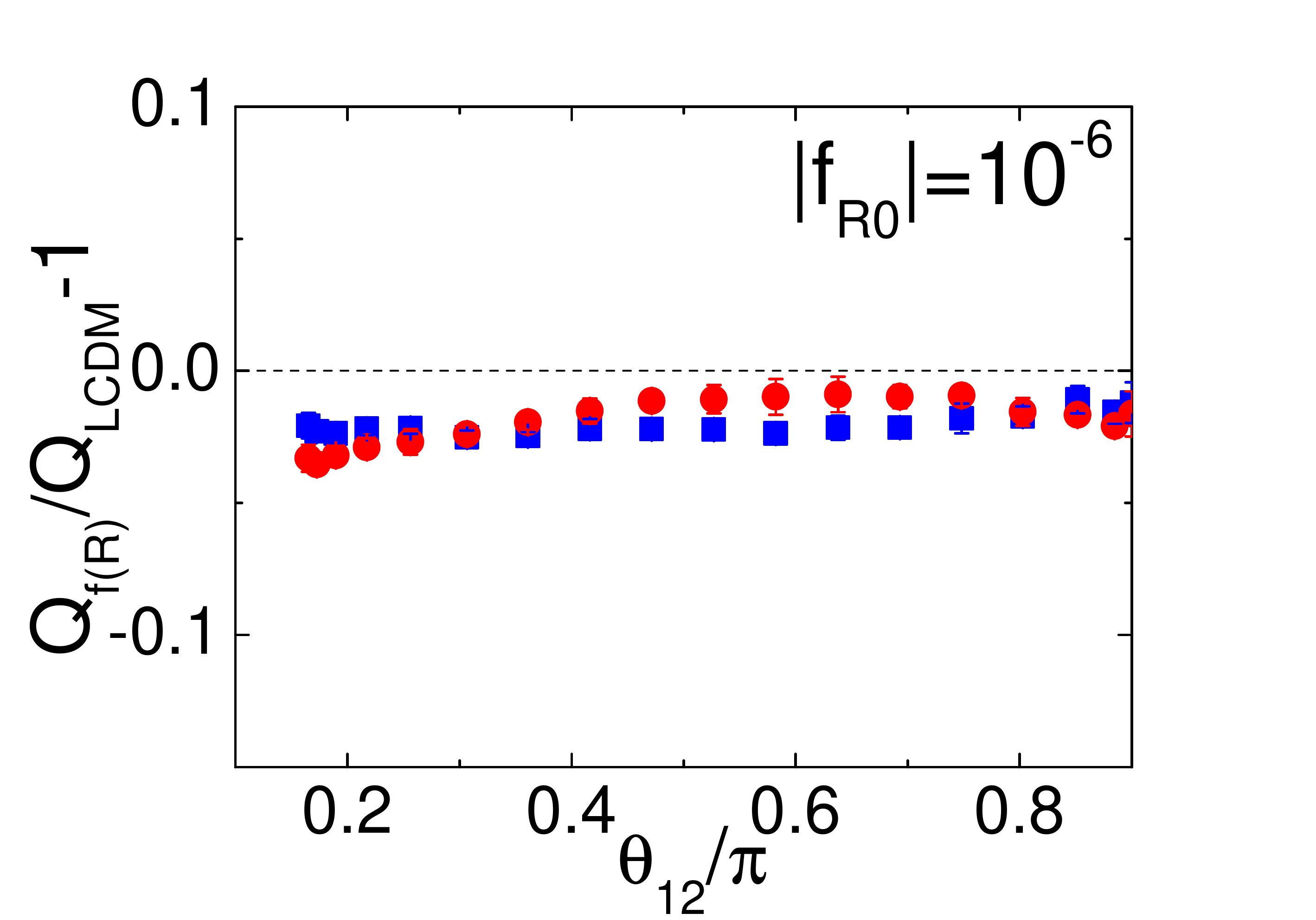}
\includegraphics[scale=0.2]{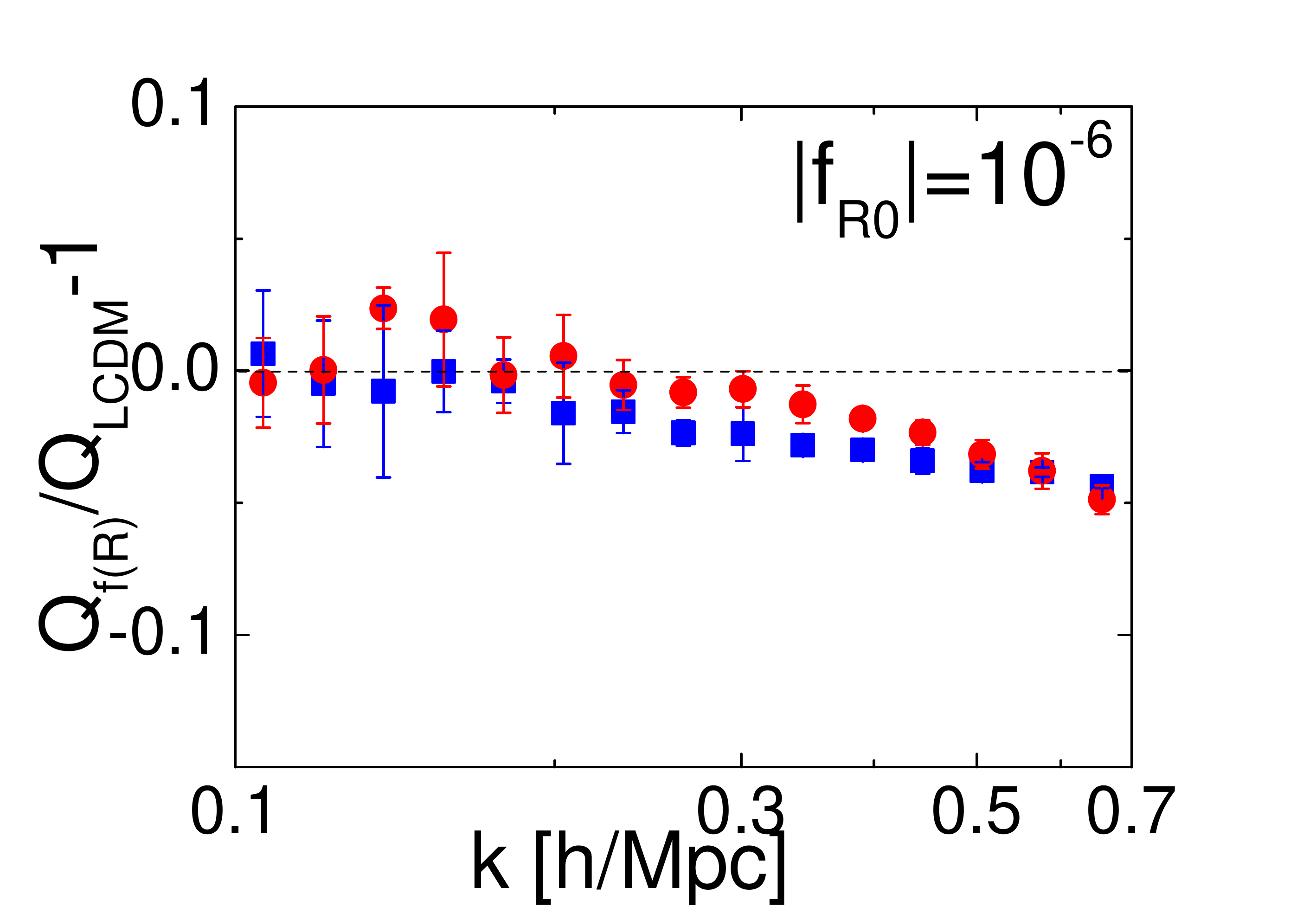}

\caption{Relative dark matter reduced bispectrum deviations (following method A) between $\Lambda$CDM and $f(R)$ models for $k_2=2k_1=0.4$ $h$/Mpc (left panels) and equilateral configuration (right panels) at $z=0$ as a function of the angle between ${\bf k}_1$ and ${\bf k}_2$, namely $\theta_{12}$ (left panel) and as a function of $k$ (right panel) for $|f_{R0}|=10^{-4}$, $10^{-5}$, $10^{-6}$ (top to bottom).  Blue points (squares) correspond to chameleon simulations and red points (circles) to non-chameleon.  Both $\Lambda$CDM and $f(R)$ bispectra have been computed from N-body simulations with the same initial conditions. As a consequence, the corresponding final ($z=0$) power spectra of the compared models are different. Error bars are the 1-$\sigma$ standard deviation of the ratio of $Q$ values amongst the 6 independent runs. Because of that, the errors due to cosmic variance cancel out. Only $L=400\, \mbox{Mpc}/h$ side-box runs are used.}

\label{bis_extra}
\end{figure}

In Fig. \ref{bis_extra} we show the dark matter reduced bispectra deviation between $\Lambda$CDM and $f(R)$ for $k_2=2k_1=0.4 h/$Mpc (left panels) and for equilateral configurations (right panels) for $z=0$ according to method A. The top panels correspond to $f(R)$ theories with $|f_{R0}|=10^{-4}$; $|f_{R0}|=10^{-5}$ for middle panels; and $|f_{R0}|=10^{-6}$ for bottom panels. The blue points correspond to full $f(R)$ theories whereas the red points to non-chameleon ones. Deviations of $f(R)$ bispectra with $|f_{R0}|=10^{-4}$ with respect to $\Lambda$CDM present a characteristic shape dependence, where the difference is maximal for $\theta_{12}\sim 0$ and $\pi$, and minimal for $\theta_{12}\sim 0.6\pi$, for both chameleon and non-chameleon and it increases as the scale is reduced. A similar trend is present for non-chameleon theories with $|f_{R0}|=10^{-5}$. On the other hand, chameleon theories with $|f_{R0}|=10^{-5}$ present a constant deviation from $\Lambda$CDM of $\sim10\%$. For $|f_{R0}|=10^{-6}$, both chameleon and non-chameleon present a constant deviation from $\Lambda$CDM of $\lesssim5\%$ and is consistent with 0.  The errors of Fig. \ref{bis_extra} are suppressed compared to the individual cosmic variance errors because we are taking the ratio of N-body simulations with the same initial power spectrum
and phases.
Therefore, we can conclude that having the same initial conditions,   the dark matter bispectra of $\Lambda$CDM and $f(R)$ theories is significantly different for $|f_{R0}|\gtrsim10^{-5}$, especially for elongated triangles ($\theta_{12}\simeq 0,\pi$) and can reach deviations in the reduced bispectra up to $\sim10\%$ for $|f_{R0}|=10^{-5}$ and up to $\sim12\%$ for $|f_{R0}|=10^{-4}$.  
We see a similar dependence on triangle shape as shown in Fig.~5 of \cite{bernardeau} (note that $\beta$ as defined there is $1/\sqrt{6}$ for $f(R)$).  

Although differences in the final dark matter power spectra  between $\Lambda$CDM and $f(R)$ theories of the same initial power are large and potentially easier to test than those in the bispectra, it is possible that the galaxy power spectra for $\Lambda$CDM and $f(R)$ models could still be similar for some particular galaxy bias model \citep{song}. Since the galaxy bias acts differently on the power spectrum and on the bispectrum, it would be very unlikely that the same galaxy bias could make $P_{\rm gal}^{\lcdm}=P_{\rm gal}^{f(R)}$ and $Q_{\rm gal}^{\lcdm}=Q_{\rm gal}^{f(R)}$ simultaneously.

Conversely, changes in the initial power spectra between the models might conspire to
make an $f(R)$ model look like a $\Lambda$CDM model for the power spectrum at $z=0$.
These can be hidden from the CMB at high redshift if they only occur at high $k$.
Because of that, in the next section we assess the differences between the $\Lambda$CDM and $f(R)$ dark matter reduced bispectra in the case where both models  have the same final power spectrum.

\subsection{Method B (matched final power spectrum)}\label{methodB}

The final power spectra of the $f(R)$ models deviate significantly from that of the $\Lambda$CDM model with the same initial conditions (see 
Fig. 2 of \cite{II}).  
These deviations reach $\sim50\%$ for $|f_{R0}|=10^{-4}$; $\sim10\%$ for $|f_{R0}|=10^{-6}$; all at $k\simeq1$ $h$/Mpc and at $z=0$.  
That begs the question of whether bispectrum differences seen in Method A are driven by these final power spectrum differences or by uniquely gravitational modifications.

To address this question, we would like 
 to adjust the initial conditions of the $f(R)$ simulations until the final power spectra match that of $\Lambda$CDM
at $z=0$.  However, the $f(R)$ simulations are computationally very expensive
(a factor of $\sim$20 increase over ordinary GR simulations).   Instead we do the converse:
we adjust the initial conditions of the 
 $\Lambda$CDM model until its final power spectrum matches the $f(R)$ simulations.
 Matching $\Lambda$CDM to the $f(R)$ simulations still tests whether the remaining 
 bispectra difference 
between the two models reflects  gravitational modifications, independently of
 power spectrum differences.

\subsubsection{Power Spectra  Matching}

In order to match final power spectra  we need a means of
quickly predicting the impact of adjusting initial conditions in $\lcdm$.  HaloFit  \citep{smith03} provides
an approximate analytic mapping between the initial and final power spectra. 
Using HaloFit we can determine the desired initial conditions, run the matching 
$\lcdm$ simulations and compare the bispectra with those of the $f(R)$ simulations.\footnote{An alternative (much faster) approach would be, instead of re-running N-body simulations and computing the evolved bispectrum from there,  to use fitting formulae from the literature to predict the $\lcdm$ bispectrum from the matched power spectrum which add a small running of the tilt. We have attempted this route; unfortunately we have found that available fitting formulae are not  sufficiently accurate  for our purposes \citep{HGMinprep}; as we will see we need a relative accuracy of better than 4\% on the reduced bispectrum which exceeds that of available fitting formulae in the mildly non-linear regime of interest.}

\begin{figure}
 \centering
\includegraphics[scale=0.25]{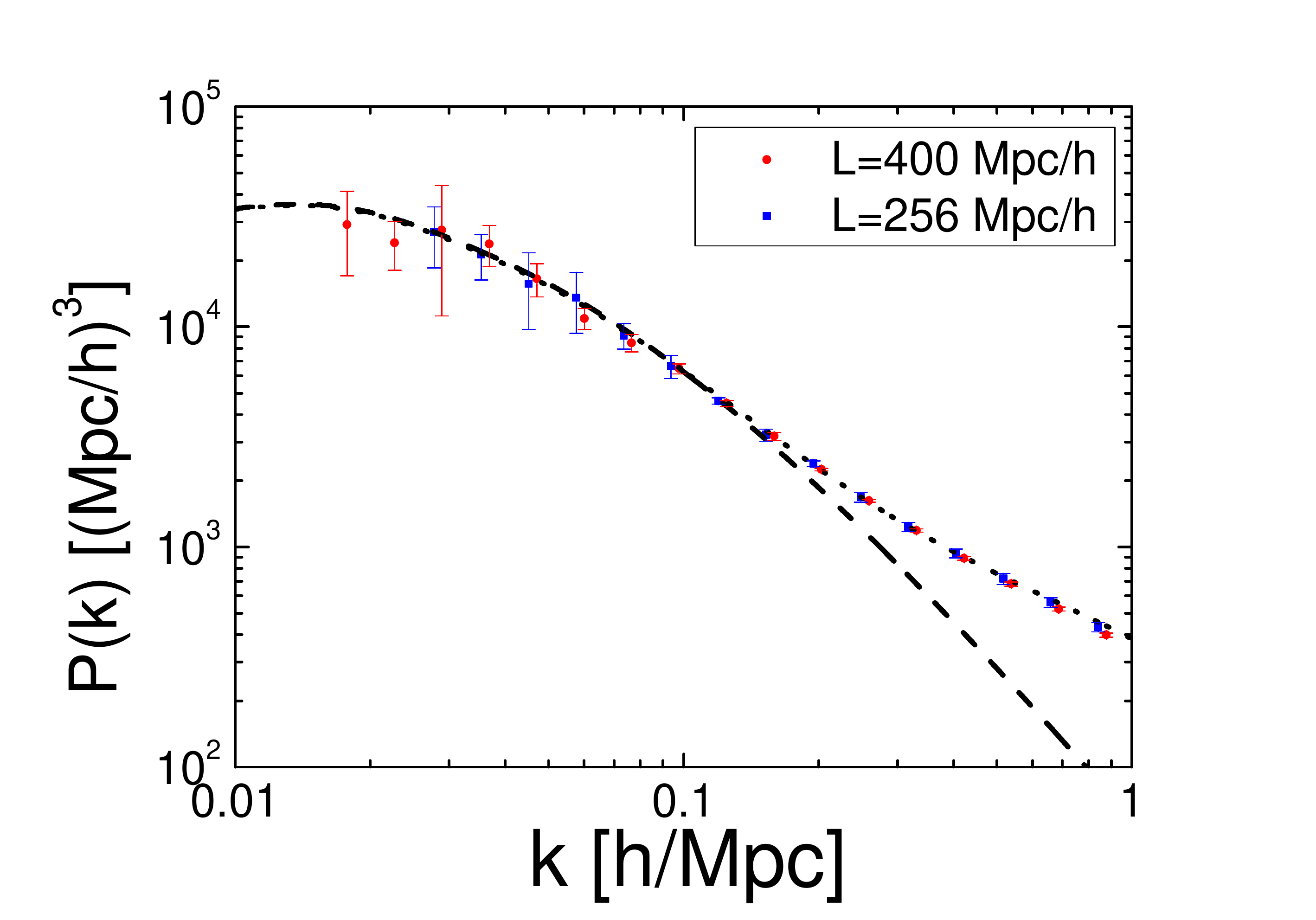}
\includegraphics[scale=0.25]{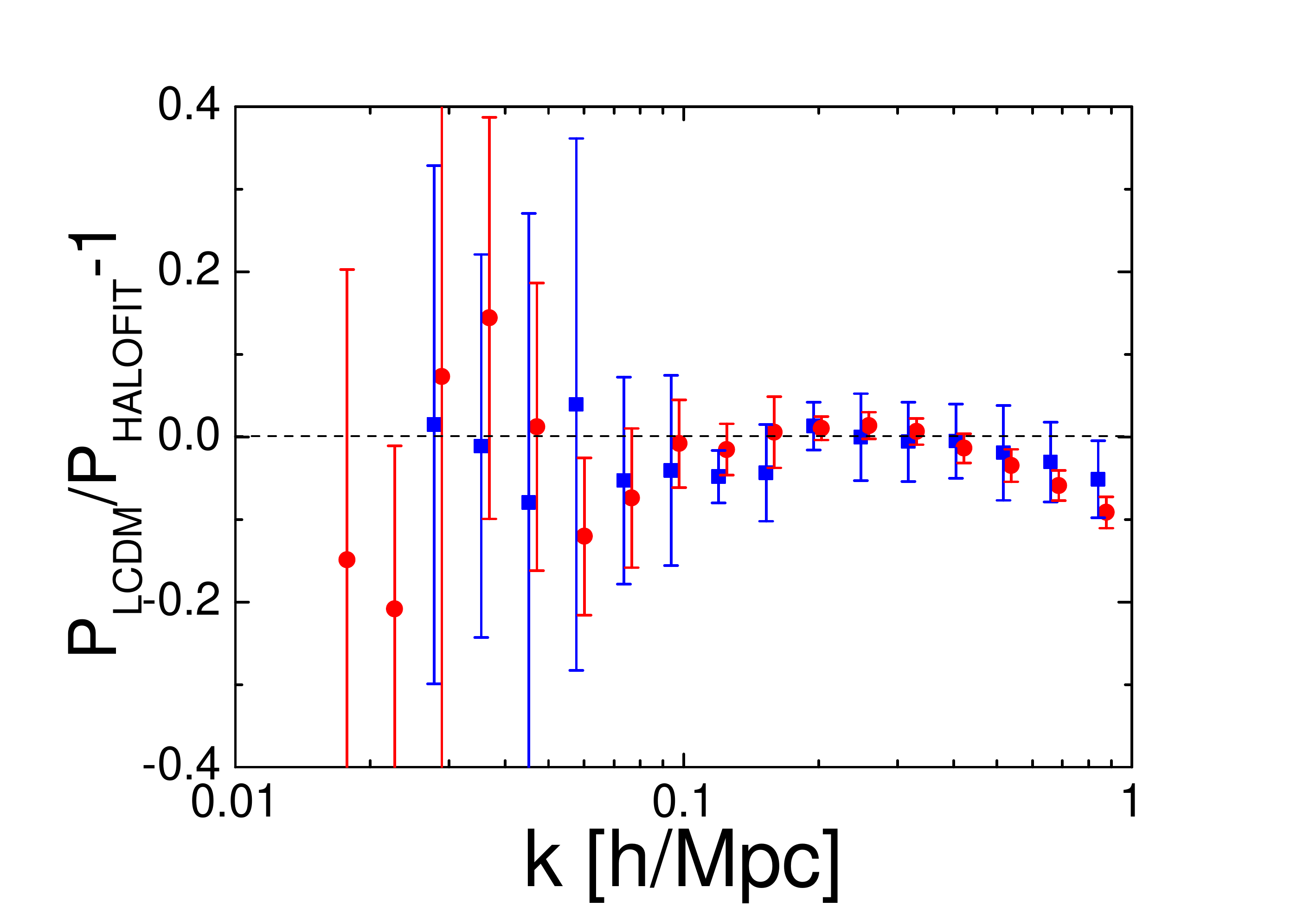}

\caption{HaloFit accuracy for the power law $\lcdm$ simulations
with box size  $L=256$ Mpc/$h$ (blue squares) and $L=400$ Mpc/$h$ box size (red circles)
{\itshape Left panel}:   linear power spectrum (dashed line) and 
non-linear prediction from HaloFit (dotted line) plotted with the simulation results. {\itshape Right panel}: fractional difference of simulation results and the HaloFit prediction.
In all cases the errors correspond to 1-$\sigma$ standard deviation amongst 6 independent runs.}
\label{power_spectra}
\end{figure}

We first test the accuracy of HaloFit in modeling the $\Lambda$CDM simulation results
(see Fig.~\ref{power_spectra}).
 In the HaloFit computation we use the same transfer function as employed in the ENZO code. We see that for the $L=400$ Mpc/$h$ runs the data-points with $k>k_N/2\simeq0.50$ $h$/Mpc underestimate the value of the predicted power spectrum by HaloFit. Here, $k_N$ is the Nyquist mode defined as $k_N=\pi N_p^{1/3}/(2L)$. We can in principle solve this limitation by using smaller boxes. For the $L=256$ Mpc/$h$ runs, $k_N/2\simeq0.79\, h\mbox{/Mpc}$ and we see that up to this scale the simulation agrees with the theoretical prediction. However the errors increase considerably as we reduce the box-size.  Error 
 bars in Fig.~\ref{power_spectra} correspond to 1-$\sigma$ standard deviation amongst  6 independent runs.  Likewise at low $k$, the simulations carry large sampling errors even for the largest boxes. To evade these problems,
 we use HaloFit to model only relative differences between simulations of the same 
$L=400$ Mpc/$h$  size, resolution and initial phases as we shall now describe.

In order to match the excess small scale final power in the  $f(R)$ model, we add an extra running of the spectral tilt parameter to the $\Lambda$CDM initial power spectrum.
Specifically, we  assume a 3 free-parameter initial power spectrum model: 
\begin{equation}
P_i(k)=P_0 \left(\frac{k}{k_p}\right)^{n_0+\frac{1}{2}\alpha\ln(k/k_p)},
\end{equation}
where $P_0$ is the amplitude of the power spectrum    at $k_p = 0.1 h$Mpc$^{-1}$  and
 $z=0$ without the effect of the transfer function and $\alpha$ is the running of the tilt,
\begin{equation}
\frac{d\ln P_i}{d\ln k} = n_0 + \alpha\ln(k/k_p).
\end{equation}
We therefore have 3 parameters ${\bf p}= \{ P_0, n_0, \alpha \}$ which specify the initial conditions.  

To find the best-fitting three parameters for a given model, we take the 
simulation results for the power spectrum ratio (following method A),
\begin{equation}
R_{\rm sim}(k) \equiv \left\langle {P^{\rm sim}_{f(R)}(k) \over P^{\rm sim}_{\Lambda \rm CDM}(k) }\right\rangle.
\end{equation}
Next we use the HaloFit prescription $P^{\rm HF}(k; {\bf p}^{\rm match})$ for the non-linear matter power spectrum at $z=0$ to find the best parameter set 
${\bf p}^{\rm match}$, by minimizing the $\chi^2$ given by
\begin{equation}
\chi^2 = \sum_j \frac{1}{\sigma^2_{R_{\rm sim}}(k_j)} \left[ 
\frac{P_{\rm  HaloFit}(k_j;{\bf p}^{\rm match})}{P_{\rm HaloFit}(k_j;{\bf p}^{\rm 0})} - R_{\rm sim}(k_j)\right]^2,
\label{eq2}
\end{equation}
where the sum runs over bins in $k$.  Here, ${\bf p}^0$ describes
the initial power spectrum used for the $f(R)$ simulations (see first
line in Tab.~\ref{tabla}).  
Finally we simulate  a matched $\Lambda$CDM simulation with the same initial phases
as the original but with a rescaling of the initial power
\begin{equation}
P_{\rm IC}(k;{\bf p}^{\rm match}) = P_{\rm IC,orig.}(k) { P_{i}(k;{\bf p}^{\rm match}) \over 
P_i(k;{\bf p}^{\rm 0})}.
\end{equation}
In order to avoid confusion, we designate these $\Lambda$CDM simulations as ``matched''; whereas $\Lambda$CDM without this modifier denotes the standard, power law, initial conditions (the one used in \S \ref{methodA}).

The advantage of this matching method is that we only model relative deviations with the
HaloFit prescription.   Thus the cosmic variance of the original simulations scale out as
do absolute errors in HaloFit, initial condition generators, resolution, etc. 
In Table \ref{tabla} we show the best-fit values of the 3 initial power spectrum parameters.
We have only used $R_{\rm sim}(k\leq0.5\:h/\rm Mpc)$ for the minimization.

\begin{table}
\begin{center}

\begin{tabular}{cccccc}
& $|f_{R0}|$ & $P_0/[10^3 (\mbox{Mpc} h^{-1})^3]$ & $n_0$ & $\alpha$ & $\sigma_8$ \\ 
\hline
${\bf p}^0$ & 0 & 6.31 & 0.958 & 0 & 0.824 \\ 
\hline
& $10^{-4}$ & 7.60 & 1.061 & -0.00643 & 0.944 \\ 
cham. & $10^{-5}$ & 6.62  & 1.087 & 0.0427 & 0.878 \\ 
& $10^{-6}$ & 6.34 & 0.985 & 0.00964 & 0.834 \\ 
\hline
non& $10^{-4}$ & 7.66 & 1.059 & -0.00963 & 0.947 \\ 
cham.& $10^{-5}$ & 6.72 & 1.141 & 0.0620 & 0.898 \\ 
& $10^{-6}$ & 6.37 & 1.012 & 0.0201 & 0.843
\end{tabular}
\end{center}
\caption{Best-fit linear $\lcdm$ initial power spectrum parameters $P_0$, $n_0$, $\alpha$ (see text) that match the N-body power spectrum at $z=0$ for each $f(R)$ simulation. These values have been computed minimizing the $\chi^2$ of the N-body non-linear power spectrum and the non-linear HaloFit power spectrum for $k\leq k_N/2\simeq0.5$ $h$/Mpc. The same transfer function and the same cosmology is used in all cases. Also $\sigma_8$ is shown for clarity.}
\label{tabla}
\end{table}

 \begin{figure}
 \centering
\includegraphics[width=0.6\textwidth]{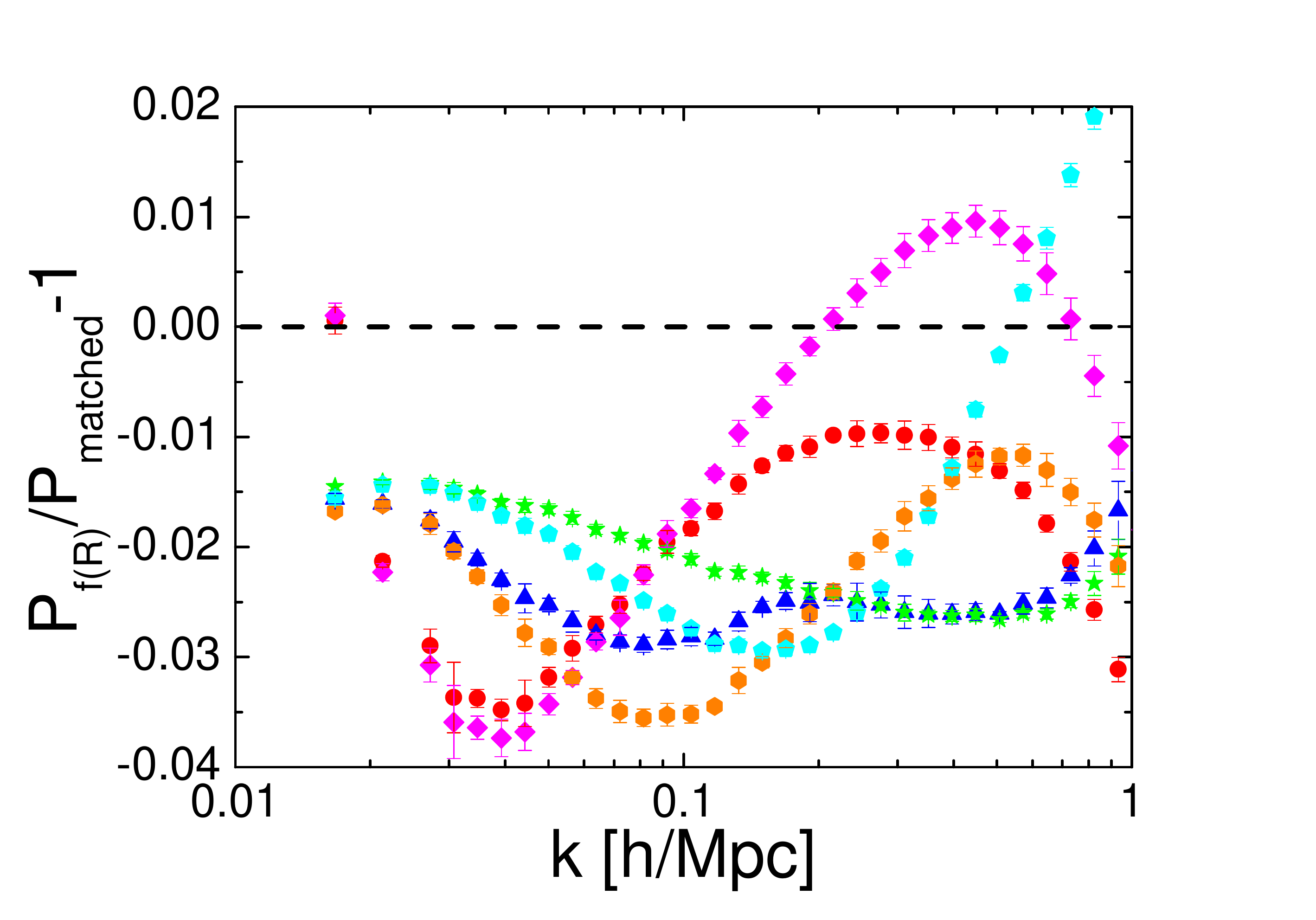}
\caption{Relative power spectrum offset between $f(R)$ simulations and corresponding matched-$\Lambda$CDM simulations for different $|f_{R0}|$ values: for the chameleon simulations with $|f_{R0}|=10^{-4}$ (red circles), $10^{-5}$ (blue triangles), $10^{-6}$ (green stars); and for the non-chameleon simulations with $|f_{R0}|=10^{-4}$ (pink diamonds), $10^{-5}$ (orange hexagons), $10^{-6}$ (cyan pentagons). The matched simulations make use of initial power spectrum conditions shown in Table \ref{tabla} found by fitting relative deviations with HaloFit.  }
\label{HaloFit_check}
\end{figure}

In Fig. \ref{HaloFit_check} we show $P_{f(R)}/P_{\matched}-1$, where $P_{f(R)}$ is the power spectrum of the $f(R)$ simulations as before and $P_{\matched}$ is the power spectrum of the matched $\Lambda$CDM simulations. Fig. \ref{HaloFit_check} indicates that HaloFit is an excellent tool to predict relative differences in non-linear power spectra even for (some) non-standard $\Lambda$CDM models. We see that the differences between the matched $\Lambda$CDM and $f(R)$ power spectra are up to $\sim4\%$ in the range $0.1 h/\mbox{Mpc}< k < 1 h/\mbox{Mpc}$, although for most of the scales and the cases are about 2-3\%. 

\begin{figure}
 \centering
\includegraphics[width=0.6\textwidth]{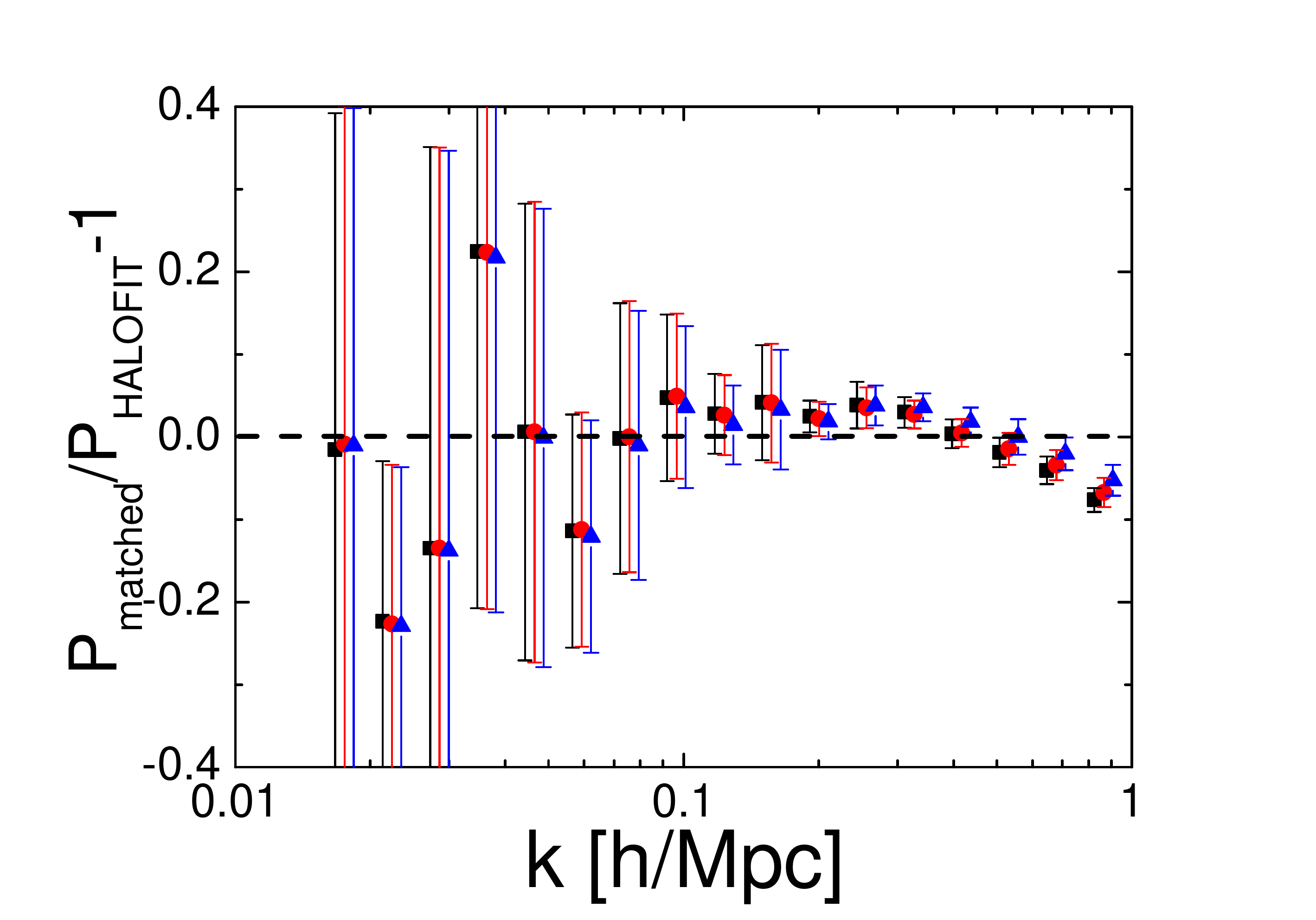}
\caption{ 
HaloFit accuracy for a representative set of matched $\Lambda$CDM models
(matched to chameleon $|f_{R0}|=10^{-4}$, black squares;  $10^{-5}$ red circles; $10^{-5}$, blue triangles).   
HaloFit is accurate within the errors for  all $k < 0.5$Mpc/$h$ for these models which contain 
running of the tilt.  Deviations from sample variance and resolution in the simulation seen here  are largely absent in the relative matching technique shown in Fig.~\ref{HaloFit_check}}.
\label{deviations2}
\end{figure}

As an aside, we can also test the absolute accuracy of HaloFit's prediction for the power spectra of the matched models.    Examples for different matched models  are shown in Fig. \ref{deviations2}.  HaloFit produces a good fit compared with the
sample variance errors for all $k < 0.5$Mpc/$h$.  As in the pure $\Lambda$CDM case,  the sample variance at low $k$ in the simulations is quite large.  Deviations up to $10\%$ for $k<1.0$ Mpc/$h$ likewise appear due to the limited simulation resolution.  Our modeling of relative effects eliminates these small differences.

In reality one does not observe at a single $z$ but in a wide z-range. As mentioned above it is not possible to  match the power spectrum at widely separated redshifts  simultaneously and this  feature can provide observational signatures independent from the bispectrum. We  can quantify this further by estimating over what redshift interval the power spectrum matching is expected to hold. Changes in the $P_{f(R)}/P_{\Lambda \rm CDM}$ were studied in  detail by \cite{II} and the excess evolves on the Hubble time scale.  Therefore we
generically expect that the matching evolves across a redshift interval of $\Delta z=1$, i.e. no faster than any other aspect of the modeling.%Changes in the $P_{f(R)}/P_{LCDM}$ were studied in  detail by \cite{II}. Assuming linear theory we can see how the minimum redshift separation for  the simultaneous spectrum matching  to fail,  increases  with decreasing $|f_{R0}|$. In fact for $k=0.2$ $h$/Mpc,  and $|f_{R0}|=10^{-4}$ the matching is up to 6\% at $z=0.3$; for $|f_{R0}|=10^{-5}$ the matching reaches 5\% deviation at $z=0.37$; whereas for $|f_{R0}|=10^{-6}$ the matching is about 3\% at $z=1.22$.

\subsubsection{Bispectrum}\label{subsection_bispectrum}

With the simulations of the matched $\Lambda$CDM models,
we  can now compare the bispectra for $\Lambda$CDM and $f(R)$ models
whose final power spectra match to a few percent.

\begin{figure}
 \centering
\includegraphics[scale=0.2]{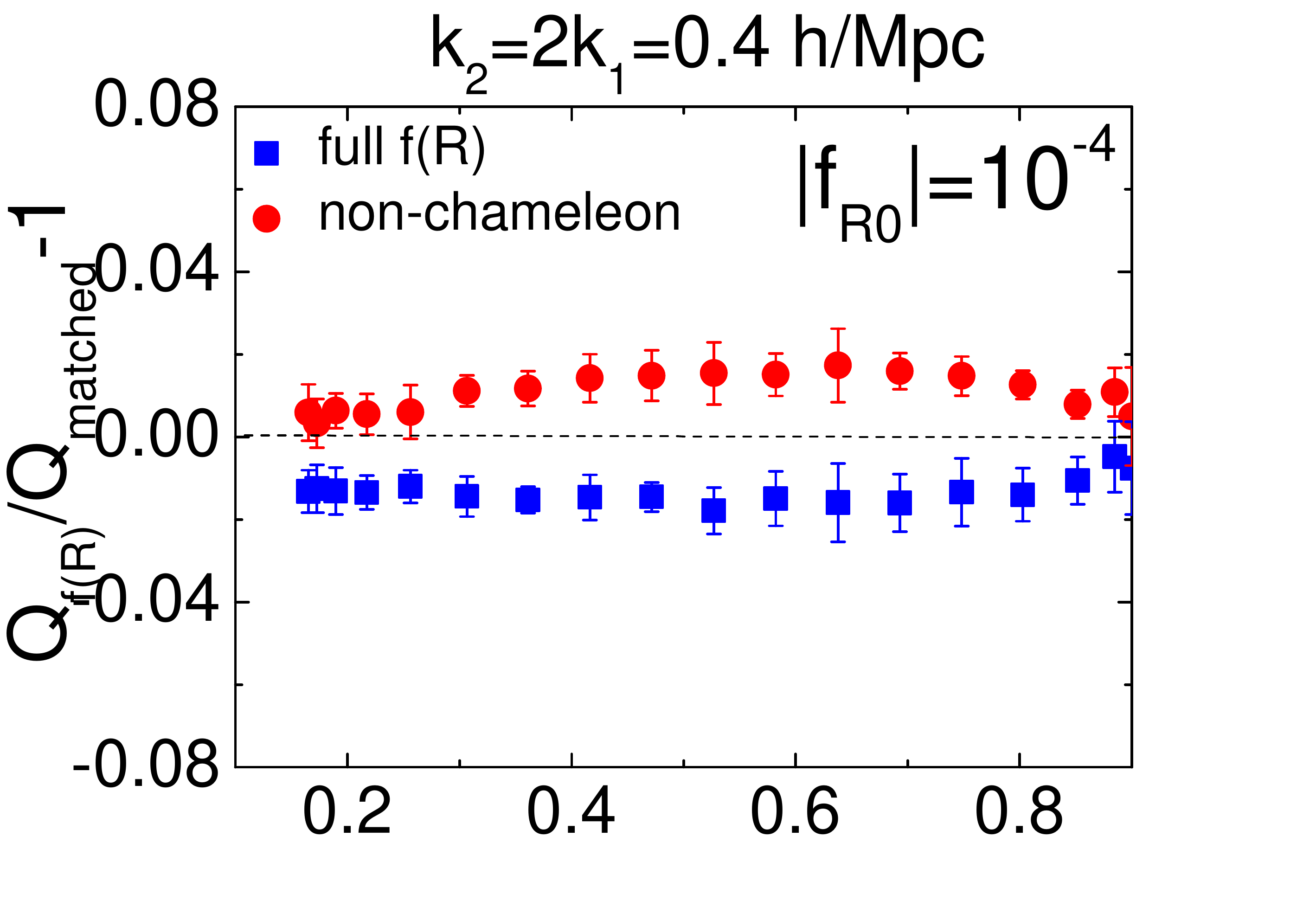} 
\includegraphics[scale=0.2]{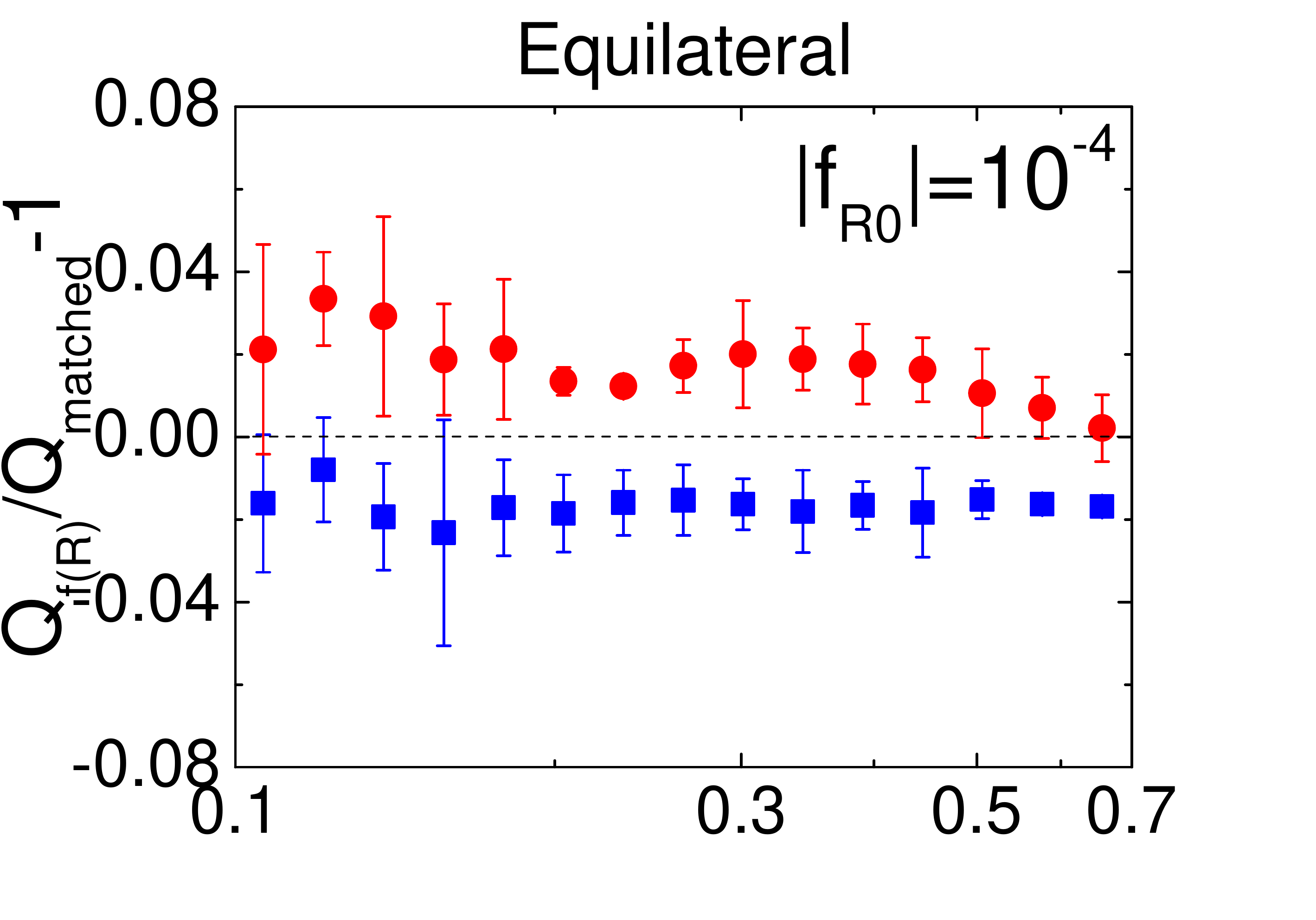}

\includegraphics[scale=0.2]{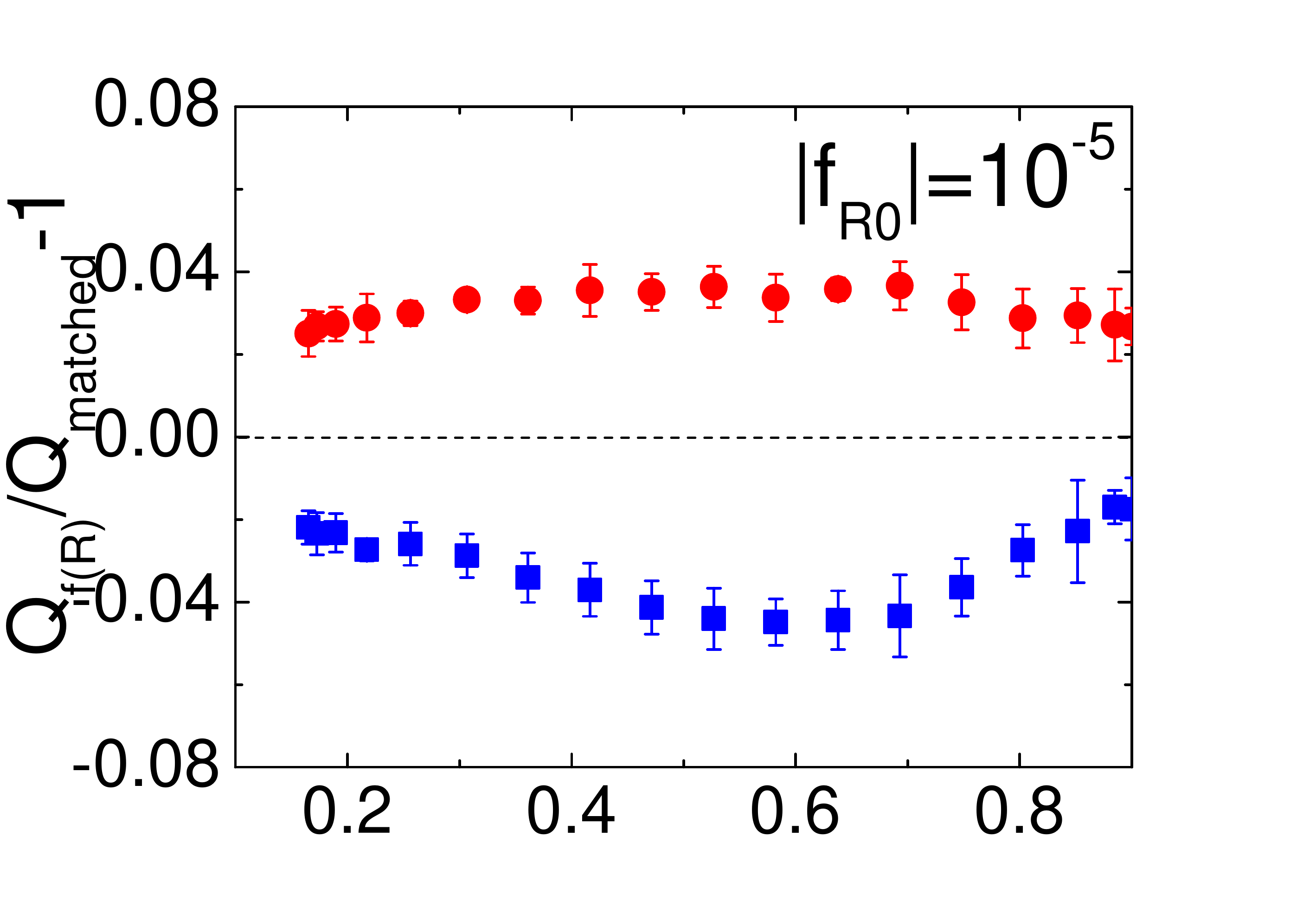} 
\includegraphics[scale=0.2]{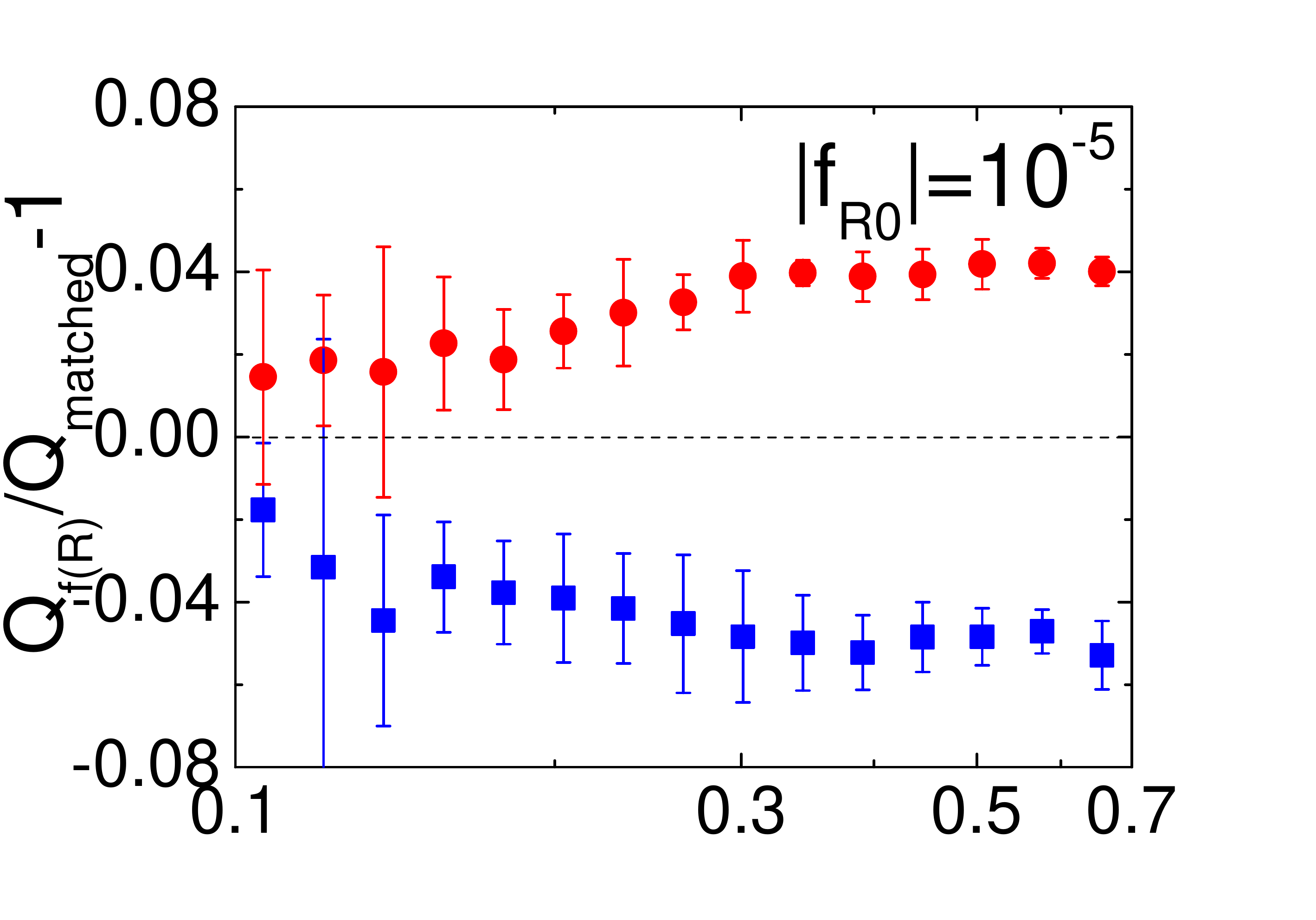}

\includegraphics[scale=0.2]{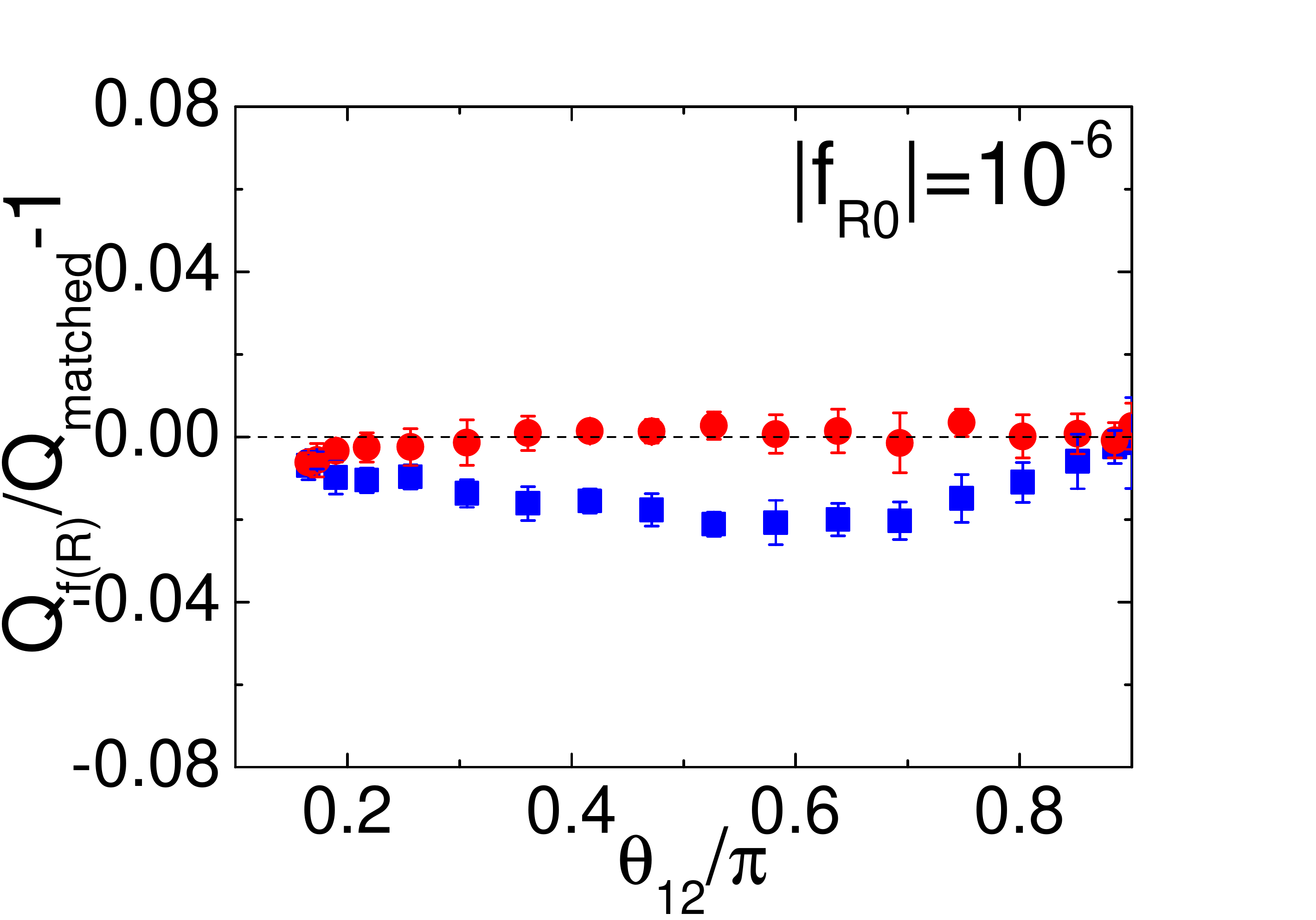} 
\includegraphics[scale=0.2]{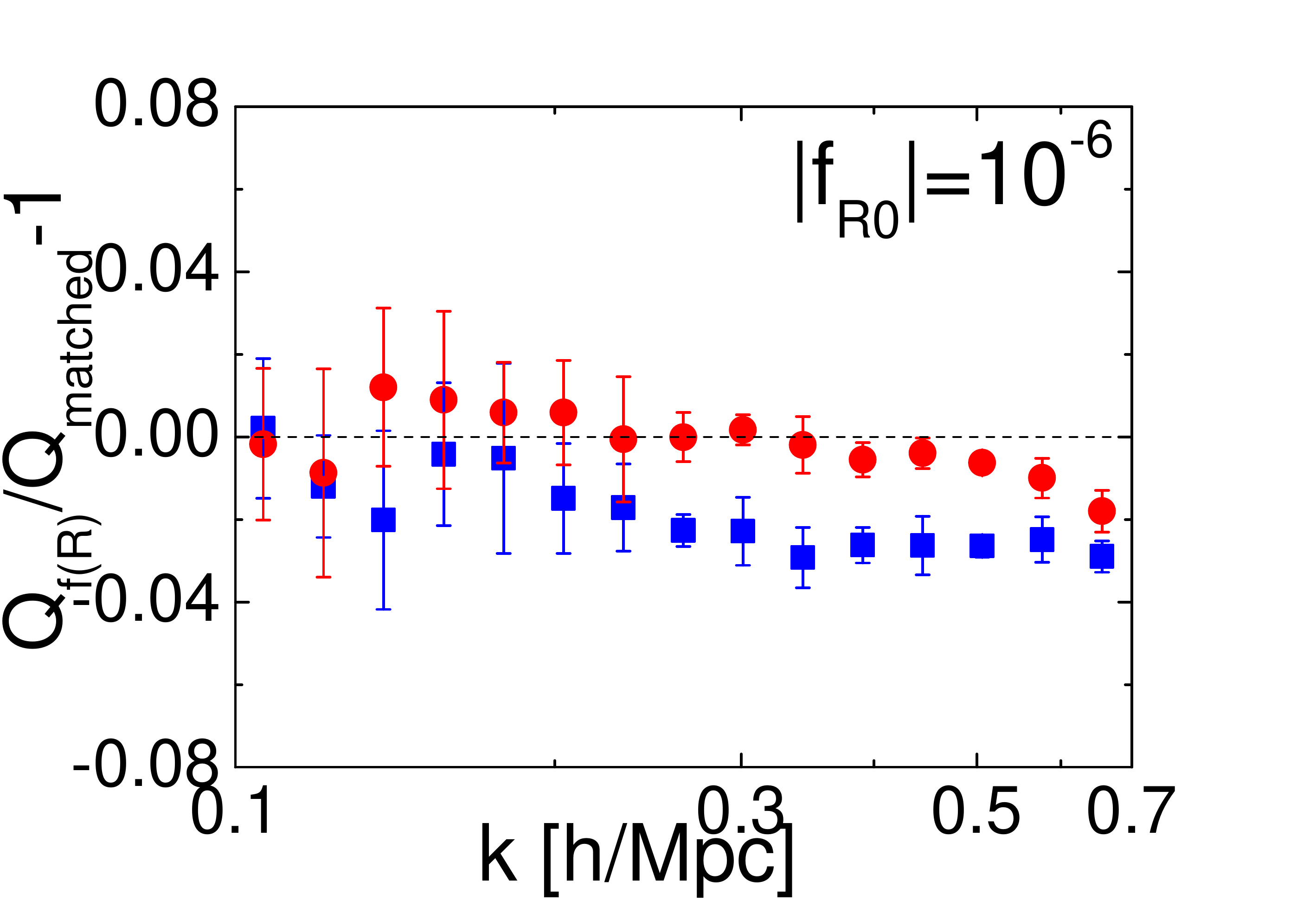}
\caption{Relative reduced bispectrum deviation for matched final power spectra (method B)  between $f(R)$ and  $\Lambda$CDM simulations for $k_2=2k_1=0.4\, h/\mbox{Mpc}$ as a function of $\theta_{12}$ (left panels), and equilateral configuration as a function of $k$ (right panels), all at $z=0$. Upper panels correspond to $|f_{R0}|=10^{-4}$, middle panels to $|f_{R0}|=10^{-5}$ and bottom panels to $|f_{R0}|=10^{-6}$. Red points correspond to non-chameleon simulations, whereas blue points to full $f(R)$ simulations. The error-bars are the $1\sigma$ standard deviation amongst the ratio of 6 independent runs. Since we are taking the ratio between runs with the same initial phases, the cosmic variance errors are not present.}
\label{bisC1}
\end{figure}

In Fig. \ref{bisC1} we show $Q_{f(R)}(k)/Q_{\matched}-1$ for $k_2=2k_1=0.4\, h/\mbox{Mpc}$ (left panel) and for equilateral triangle configuration (right panel), where $Q_{f(R)}(k)$ is the reduced bispectrum for $f(R)$ simulations, and $Q_{\matched}$ is the reduced bispectrum for the matched $\Lambda$CDM simulations. Red points show the ratio for non-chameleon simulations whereas blue points for chameleons ones. Top panels correspond to $|f_{R0}|=10^{-4}$, middle panels to $|f_{R0}|=10^{-5}$ and bottom panels to $|f_{R0}|=10^{-6}$.
In particular, we see that for the chameleon and non-chameleon cases with $|f_{R0}|=10^{-4}$ and $10^{-6}$ the deviation is very close to 0 $(\lesssim 2\%)$.
 For the $|f_{R0}|=10^{-5}$ some differences appear: for the non-chameleon case there is an excess of $\sim4\%$ and for the chameleon case there is a deficit of $\sim 4\%$ in $Q_{f(R)}$ respect to $Q_{\matched}$, both within 5-6$\sigma$.  The value $|f_{R0}| \sim 10^{-5}$ is special in that it marks the onset of the chameleon mechanism in the largest structures 
 in the simulations.   
The chameleon effect may have a small but measurable impact on $Q$ in this
  transition region where the chameleon effect is present for some but not all structures.
  Analogous transient enhancements appear in the mass function \citep{LiHu11}.    One
  should  bear in mind though that this difference is of order the difference in the matched power 
  spectra which varies between the full and no-chameleon cases.

Thus, for all values of $f_{R0}$ deviations are below $\sim 4\%$. In particular we do not observe that squeezed triangles (those with $\theta_{12}\simeq 0, \pi$) present higher deviations between different gravity models as has been observed in method A (Fig. \ref{bis_extra}) and predicted from theoretical models that followed the same assumptions as adopted in method A \citep{borisov,bernardeau}.

Finally, we found that it is better to analyze the deviation between reduced bispectra $Q$ rather than between bispectra $B$. This is because  the power spectrum dependence is partially canceled in the reduced bispectra. In spite of having run $\Lambda$CDM simulations to match the $f(R)$ power spectra, several percent differences between $f(R)$ and matched $\Lambda$CDM power spectra are still present (Fig.~\ref{HaloFit_check}).  These lead to higher deviations in $B$ between the models (up to $\sim 8\%$ in some cases) than in $Q$.  Thus, using $Q$ instead of $B$ is much more robust if we want to compare models with similar power spectra. Of course, one should keep in mind that not all the $P(k)$-dependence is cancelled when using $Q$, as evidenced by comparing with the results of method A.   Strictly speaking, this is only true for equilateral configurations and up to tree-level in Eulerian perturbation theory.

Finally, one may want to make a connection between these results and some analytic model, namely perturbation theory (PT). Since at tree level in PT the reduced bispectrum is independent of the power spectrum (at least for equilateral configuration), the differences observed between Fig. \ref{bis_extra} and \ref{bisC1} should be due to higher order corrections in $\Lambda$CDM. At 1-loop,  corrections to the bispectrum  can be found in e.g., \cite{Sefusatti09,SC97}. One can see that  the leading terms depend on the linear and one-loop power spectrum and weakly on cosmology and gravity through the standard tree level bispectrum kernel (see \cite{bernardeau} for a modification of this kernel for some $f(R)$ theories). Thus, a small modification of this formula could be expected due to $f(R)$ gravity. However, this interpretation  should be considered more in a qualitative way than in a  strictly quantitative way. In fact one should take into account that the precision of 1-loop PT for the bispectrum is not much better than the other (phenomenological) analytic formulae \citep{HGMinprep}.  As we have already mentioned, currently there is no analytic model that predicts the bispectrum at scales of interest here with an accuracy of few percent.

\subsubsection{Discussion}

If the remaining $\sim 4\%$ deviation for $|f_{R0}|=10^{-5}$ reflects gravity and not the residual mismatch in power spectra, then it is in principle measurable with large-volume surveys.
In this work, considering only the 6 runs of 400 Mpc/$h$ box-size and provided that $h=0.73$, the total volume is $6\times( 0.4\, \mbox{Gpc}/h)^3\simeq1 \mbox{Gpc}^3$. 
We expect that future surveys will cover larger volumes: BOSS\footnote{Baryon Oscillation Spectroscopic Survey} $V\sim5\, (\mbox{Gpc}/h)^3$, DES\footnote{Dark Energy Survey} $V\sim 10\, (\mbox{Gpc/h})^3$ or EUCLID $V\sim 100\, (\mbox{Gpc}/h)^3$.  As the 6 runs have different initial conditions we can use them to estimate the  expected error on $Q$ in the limit that it is dominated by cosmic variance. We have measured that the  error in $Q$ for our simulations at scales of $k \sim 0.3 h/\mbox{Mpc}$ is about $5\%$. We assume that the variance scales as the inverse of the number of modes, and thus the standard deviation approximately scales as $V^{1/2}$ .
Therefore, for a 10 $\mbox{Gpc}^3$ survey the error bars could, in principle, be as much as  $\sqrt{10} \sim 3$ times smaller than our prediction.
This implies that a survey with $>10\,\mbox{Gpc}^3$ volume  (e.g., DES, EUCLID) would yield an error on $Q \sim 2\%$ at these scales.  Since the expected deviation may be of order 4\%, having smaller errors would help us to confirm or discard possible deviation of the bispectrum due to modifications of gravity.

On the other hand, we have analyzed the dark matter bispectrum which is not  directly observable. In practice,  sources of error  that we have neglected here may appear: i) galaxy-surveys provide a biased information about dark matter, ii)   additional effects such as redshift distortions change the observed bispectrum (in fact we expect modified gravity to affect redshift distortions more strongly than the density field itself).  Also as we go to higher $z$, we expect less deviations at a given scale.   Conversely, the matched power spectra at $z=0$ would become mismatched and provide other observable effects.

The results from Fig. \ref{bisC1} provide another important result. We have seen that  two  $f(R)$ theories of gravity with indistinguishable non-linear dark matter power spectrum,  have very similar and possibly indistinguishable dark matter bispectra. This opens up the possibility of using these two statistics to break degeneracies in the galaxy bias  in a way that is robust to the assumptions about  the true underlying gravity model.

In fact the $f(R$) effects on the power spectrum are at the 20-50\%  level.  A modification of galaxy bias achieving similar effects would likely affect the reduced bispectrum at  least  at the 10\% level (for example, a linear bias term affect the power spectrum $\propto b_1^2$ and the reduced bispectrum $\propto 1/b_1$), significantly larger than the $f(R)$ effects on the reduced bispectrum.  

\section{Conclusions}\label{conclusions}
In this work we have analyzed the deviations in the reduced bispectrum produced by a modification of gravity, specifically the $f(R)$ class of models, both with and without the chameleon mechanism. In order to do that, we make use of a suite  of $f(R)$ and $\Lambda$CDM simulations. We have proceeded in two different ways to analyze the bispectrum deviation from these simulations, methods A and B, which differ in whether the initial or
final power spectra of the two cosmologies are set equal.

Method A compares the bispectrum output of $f(R)$ and $\Lambda$CDM N-body simulations with the same initial power spectrum. Fig. \ref{bis_extra} shows the bispectrum deviation obtained using this method. We observe a considerable deviation (up to $10-15\%$) in the reduced bispectrum between these $f(R)$ models and the $\Lambda$CDM one.
 Such differences in the bispectrum could be easily detected by surveys covering volumes $> 1\,\mbox{Gpc}$ such as  e.g., the on-going BOSS survey.
 Higher deviations are seen for higher values of $|f_{R0}|$ and for squeezed triangle configurations. In this method, both $\Lambda$CDM and $f(R)$ gravity runs start from the same initial $\delta_k$ values. Because of that, the different evolution of the gravity models naturally leads to different power spectra (as was observed in \cite{II}) and also to different bispectra. This way of proceeding is equivalent to the theoretical works of \citet{bernardeau,borisov}.  In order to explain discrepancies between the matter power spectrum
in $f(R)$ and the observed galaxy power spectrum, one could invoke a
scale-dependent galaxy bias.  Since galaxy bias enters into the reduced 
galaxy bispectrum in a different way than in the power spectrum, 
bispectrum measurements can in principle close this loophole.

Alternatively, the large power spectrum differences can be eliminated by changing the shape of the initial power spectrum to instead match the final dark matter power
spectrum at $z=0$.   This is at the base of method B.  In this method,
we compute the bispectrum deviation between a $\Lambda$CDM and a $f(R)$ model, both with the same final power spectra. 
Thus, we simulate a 
$\lcdm$ model with certain initial power spectrum parameters (summarized in Table \ref{tabla}) that are adjusted to best match the $f(R)$ power spectrum at $z=0$. From the simulations outputs  we compute power spectra and bispectra.  For the power spectra, residual differences are never higher than $4\%$ in the range $0.1\, h/\mbox{Mpc}<k<1\,h/\mbox{Mpc}$.

Likewise the differences in the reduced bispectrum are also smaller in the matched comparison. For the $|f_{R0}|=10^{-4}$ and $10^{-6}$ cases, the $Q$ deviation is  consistent with 0 within $1\sigma$. For $|f_{R0}|=10^{-5}$  deviations in $Q$  at most reach the $4\%$ level with $5-6\sigma$ significance.   These deviations are potentially a signature of the onset of the chameleon mechanism in the largest structures in the Universe.  However   given that
this is the same order as the power spectrum difference it is unclear whether  these differences indicate power-spectrum-independent modified gravity   effects or that the
two power spectra are not perfectly matched.   In the former case,  larger surveys like EUCLID will allow for a measurement of the bispectrum with enough precision to  obtain a $>6\sigma$ significance, even when exactly matching the power spectra.  

On the other hand, the effect of  deviations from GR gravity on the reduced bispectrum are weak compared to those on the power spectrum (at least for the cases considered here), opening up the possibility of breaking the galaxy-bias degeneracy. In fact the  effect of galaxy  bias  is expected to  be different in the power spectrum and in the bispectrum, which is why, in the context of GR gravity, the bispectrum is used to constrain galaxy bias.  While the shape of the non-linear power spectrum seems to carry  information about the underlying gravity model, one may always argue that a  shape of the evolved power spectrum not compatible with the GR predictions could be due to biasing.   For the cases we have considered here, the dependence of the reduced bispectrum on deviations from GR is weaker than the effects of bias modifications necessary to explain the deviations in the power spectrum.  While we have only studied $f(R)$ models here, there is no apparent reason why this result should be specific to $f(R)$.  Hence, if our findings were to remain qualitatively true for other gravity modifications, this would  confirm the usefulness of employing the reduced bispectrum together with the power spectrum to constrain bias parameters.

\section{Acknowledgments}
We thank Christian Wagner for useful discussions and help with N-body simulations.
HGM is supported by a CSIC JAE grant, and  thanks the Kavli Institute for Cosmological Physics (KICP) at University of Chicago for hospitality. Part of this work stemmed from discussions at the Centro de Ciencias de Benasque Pedro Pascual.
WH is supported by the KICP under
NSF contract PHY-0114422, DOE contract DE-FG02-90ER-40560 and the Packard Foundation.
LV and RJ are supported by MICINN grant AYA2008-03531.  
LV acknowledges support from  grant FP7 ERC- IDEAS Phys.LSS 240117.  FS is
supported by the Gordon and Betty Moore Foundation at Caltech.

\bibliographystyle{mn2e}

\end{document}